\documentclass[sigconf]{acmart}
\AtBeginDocument{%
  }

\copyrightyear{2025}
\acmYear{2025}
\setcopyright{rightsretained}
\acmConference[DEBS '25]{The 19th ACM International Conference on
Distributed and Event-based Systems}{June 10--13, 2025}{Gothenburg, Sweden}
\acmBooktitle{The 19th ACM International Conference on Distributed and
Event-based Systems (DEBS '25), June 10--13, 2025, Gothenburg,
Sweden}\acmDOI{10.1145/3701717.3733230}
\acmISBN{979-8-4007-1332-3/2025/06}

\usepackage{makecell}

\begin{document}

\title{Towards Stream-Based Monitoring for EVM Networks}

\author{Emanuel Onica}
\email{emanuel.onica@uaic.ro}
\affiliation{%
  \institution{Alexandru Ioan Cuza University}
  \city{Ia\c{s}i}
  \country{Romania}
}

\author{Claudiu-Nicu B\u{a}rbieru}
\email{claudiu.barbieru@gmail.com}
\affiliation{%
  \institution{Alexandru Ioan Cuza University}
  \city{Ia\c{s}i}
  \country{Romania}
}

\author{Andrei Arusoaie}
\email{andrei.arusoaie@uaic.ro}
\affiliation{%
  \institution{Alexandru Ioan Cuza University}
  \city{Ia\c{s}i}
  \country{Romania}
}

\author{Oana-Otilia Captarencu}
\email{oana.captarencu@info.uaic.ro}
\affiliation{%
  \institution{Alexandru Ioan Cuza University}
  \city{Ia\c{s}i}
  \country{Romania}
}

\author{Ciprian Amariei}
\email{ciprian.amariei@gmail.com}
\affiliation{%
  \institution{Testable Research, Inc.}
  \city{Middletown, Delaware}
  \country{USA}
}

\renewcommand{\shortauthors}{Onica et al.}

\begin{abstract}
We believe that leveraging real-time blockchain operational data is of particular interest in the context of the current rapid expansion of rollup networks in the Ethereum ecosystem. 
Given the compatible but also competing ground that rollups offer for applications, stream-based monitoring can be of use both to developers and to EVM networks governance.
In this paper, we discuss this perspective and propose a basic monitoring pipeline.

\end{abstract}

\ccsdesc[500]{Computer systems organization~Data flow architectures}
\ccsdesc[500]{Applied computing~Event-driven architectures}

\keywords{rollups, monitoring, CEP, Ethereum, DApps}

\maketitle

\section{Introduction}
Ethereum is the leading blockchain ecosystem for decentralized applications (DApps), with over two-thirds of blockchain networks that execute smart contracts relying on the Ethereum Virtual Machine (EVM)~\cite{DAppRadar}.
Among these networks, more than 40 are EVM-based rollups. 
These horizontally scale the main Ethereum network by processing transactions that trigger smart contract execution, while committing the transaction data and rollup states to the main Ethereum network for maintaining trust guarantees. 

Smart contract execution is compatible across EVM-based networks. 
This creates a comparison ground, where real-time monitoring of common metrics in different networks can be interesting for both DApp developers and for network governance bodies. 

Several platforms emerged recently, offering functionality fit for stream analytics of blokchain data. 
Firehose~\cite{Firehose} applies a blockchain node instrumentation that allows capturing block data, stored and fed through a data flow pipeline to be served by a gRPC endpoint. 
Further, the block data can be consumed using Substreams, a dedicated indexing API. 
EVM Trackooor~\cite{Zellic} is a different solution that leverages the possibility to subscribe directly via the RPC service offered by the EVM network nodes for capturing block information.

In our work, we develop on the use case perspectives of EVM networks monitoring by leveraging frameworks capable of collecting real-time data as mentioned above.
These are fit for creating a complex event processing pipeline that can be used for tracking dynamic situations in EVM-based networks.

\section{Use Case Perspectives}

There are multiple use cases for real-time EVM network monitoring. 
We focus on some that stem from the competing nature that is inherent in networks offering similar support for smart contracts.

From a DApp developer's perspective, a first use case is selecting a cost-efficient network for a DApp's contracts.
DApp users must pay for transactions in EVM-based public networks.
Fees are calculated as a product of the transaction computational cost expressed in gas units and a price per gas unit.
The latter varies depending on the network load and can be a decision factor for selecting one network or another. 
One could argue that this choice can be made relying on historical data. 
Indeed, we agree that the utility of real-time monitoring depends in this use case on the DApp's capacity to dynamically switch between different EVM chains for operating its contracts at lower cost rates. 
We have a work in progress in this direction, but this falls outside the current paper's scope.

A second use case is DApp profiling. 
Real-time monitoring can be used to track the DApp's access patterns among different users, such as time intervals or various parameters in transactions specific to the DApp.
If an application owner has similar DApps deployed on different networks, it can derive, aggregate or compare profile patterns across networks and take dynamic decisions depending on these (e.g., enforcing restrictions in case of suspicious behavior).

From the network's governance perspective, a third use case is monitoring and dynamically adjusting operational parameters of the network. 
An example is the block capacity, which is typically set to a fixed maximum amount of gas units, limiting the number of included transactions.  
This parameter can be periodically adjusted if the network governing body considers that the network load and processing capacity indicate a need for such action. 

\begin{figure}[t]
  \vspace{-7pt}
  \centering
  \includegraphics[width=\linewidth]{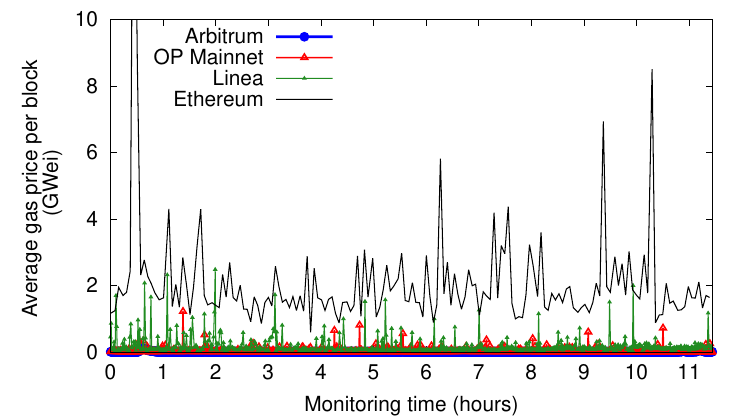}
  \vspace{-17pt}
  \caption{\label{figprice}Gas price monitored over 12 hours}
  \vspace{-9pt}
\end{figure}

\section{Basic Monitoring Architecture}

We briefly sketch a monitoring architecture layout, which we implemented to evaluate the volatility of EVM-specific metrics. 

First, we need to capture the intended metrics from the monitored networks. 
We used Trackooor~\cite{Zellic}, due to its capacity to directly retrieve blocks' information using the RPC service of the nodes, and consequently lower latency in obtaining the real-time data.

The metrics stream that we collect is further fed to Apache Kafka, as a message broker for temporary data retention purposes, which permits parallelizing subsequent stream consumers.
This is particularly useful for a complex event processing (CEP) pipeline, which might apply different transformations on similar raw metrics in the stream, or for multiple parallel monitoring pipelines.

The data stream can be further consumed by a CEP monitoring pipeline that can be formed of chained operators according to the intended use case. 
Focusing on metrics volatility, we only used an Apache Flink sink operator to collect measurements.

A specific operator, which we consider recommended in monitoring pipelines over multiple rollups, is a data normalization operator.
Despite the common transaction execution engine, EVM rollups use similar operational parameters that sometimes have different semantics. 
One example is the gas limit defining block capacity. 
Some rollups report the effective limits, while others enforce lower limits than reported. 
These rollups artificially increase reported gas limits for various reasons, e.g., Arbitrum to accommodate variations in parent chain costs~\cite{Arbitrum}, Linea to maintain a constant base fee~\cite{Linea}.

Another example of inconsistent use is the gas price on rollup networks. 
According to common EVM specifications, this is composed of a base fee and a priority fee, where both can vary. 
As mentioned above, Linea's block space management results in a constant base fee. 
Arbitrum ignores the priority fee by refunding it and considers only the base fee in the effective operation.

\begin{figure}[t]
  \vspace{-7pt}
  \centering
  \includegraphics[width=\linewidth]{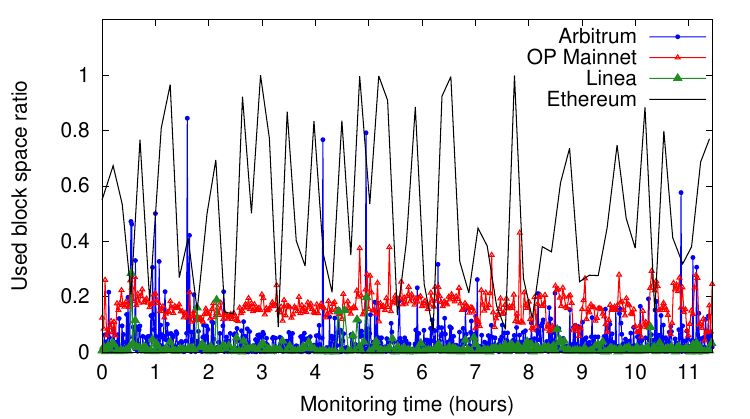}
  \vspace{-17pt}
  \caption{\label{figblock}Used block capacity monitored over 12 hours}
  \vspace{-9pt}
\end{figure}

\section{Metrics Volatility}

Among other factors, real-time monitoring in the discussed use cases is tightly related to metrics volatility. 
To test our monitoring setup, we evaluated two composed metrics, over approximately 12 hours, for three rollups: Arbitrum, OP Mainnet, Linea, and the main Ethereum network. 
We tracked the blocks during this interval, and the first metric we extracted is the gas price. 
A downsampled overview of the results is depicted in Figure~\ref{figprice}. 
The metric was relatively stable over the monitored period. 
In particular, Arbitrum maintained a low, constant base fee, which accounts for the complete price.
Ethereum was considerably more volatile than the rollups, with periodical outliers.
For the rest of the networks, the interquartile range (IQR) was relatively tight around the median. 
We provide these statistics in Table~\ref{tab:metrics}.

Second, we computed the ratio of the block space used by transactions.
Figure~\ref{figblock} displays the monitoring results. 
Again, the main Ethereum network showed considerably higher volatility than the rollups, with occasional spikes from Arbitrum. 
The IQR shows again that observed values are mostly concentrated around the median.

\begin{table}[b]
  \caption{IQR and median of gas price and used block ratio}
  \vspace{-5pt}
  \label{tab:metrics}
\begin{tabular}{|c|c|c|c|c|} 
 \hline
 Network &
 \makecell{Gas price \\ IQR} & 
 \makecell{Gas price \\ median} &
 \makecell{Block ratio \\ IQR} &
 \makecell{Block ratio \\ median} \\
 \hline\hline
Arbitrum & 0 & 0.01 & 0.044 & 0.02  \\\hline
OP Mainnet & 0.009 & 0.009 & 0.048 & 0.16  \\\hline
Linea & 0.086 & 0.092  & 0.015  & 0.009  \\\hline
Ethereum & 0.814 & 1.645 & 0.415  & 0.441 \\\hline
\end{tabular}
\vspace{-10pt}
\end{table}

\section{Conclusion}
In this paper, we discussed several use cases advocating a stream-oriented monitoring solution for EVM networks, we proposed a simple pipeline monitoring architecture, and we used it to monitor the volatility of some EVM metrics. 
We intend to expand our solution in future work, and also use alternate data provision solutions such as Firehose with Substreams for comparison purposes.  

\begin{acks}
This work was supported by a grant from the Romanian Ministry of Research, Innovation and Digitization, CNCS/CCCDI - UEFISCDI, project 86/2025 ERANET-CHISTERA-IV-SCEAL, within PNCDI IV.
\end{acks}

\bibliographystyle{ACM-Reference-Format}
\bibliography{sample-base}

\end{document}